\documentclass{article}
\usepackage{spconf,amsmath,graphicx}

\usepackage{cite}
\usepackage{amsmath,amssymb,amsfonts,amsthm}
\usepackage{bm}
\usepackage{algorithm}
\usepackage{algorithmic}

\usepackage{graphicx}
\usepackage[utf8]{inputenc}
\usepackage{subfigure}
\usepackage{textcomp}
\usepackage{xcolor}
\usepackage[unicode=true]{hyperref}
\usepackage{enumitem}
\usepackage{indentfirst}
\usepackage{multirow,multicol}

\setenumerate[1]{itemsep=0pt,partopsep=0pt,parsep=\parskip,topsep=5pt}
\setitemize[1]{itemsep=0pt,partopsep=0pt,parsep=\parskip,topsep=5pt}
\setdescription{itemsep=0pt,partopsep=0pt,parsep=\parskip,topsep=5pt}

\newtheorem{theorem}{Theorem}
\newtheorem{lemma}{Lemma}

\def\BibTeX{{\rm B\kern-.05em{\sc i\kern-.025em b}\kern-.08em
		T\kern-.1667em\lower.7ex\hbox{E}\kern-.125emX}}

\title{Sparse High-Order Portfolios via Proximal DCA and SCA}
\name{Jinxin Wang$^{\dag}$, Zengde Deng$^{\ddag}$, Taoli Zheng$^{\dag}$, and Anthony Man-Cho So$^{\dag}$}
\address{$^{\dag}$Department of Systems Engineering and Engineering Management, CUHK, Hong Kong SAR, China \\
$^{\ddag}$Cainiao Network, Hangzhou, China}
\begin{document}
%
\maketitle
\begin{abstract}
In this paper, we study the cardinality constrained mean-variance-skewness-kurtosis (MVSKC) model for sparse high-order portfolio optimization. The MVSKC model is computationally challenging, as the objective function is non-convex and the cardinality constraint is discontinuous. Since the cardinality constraint has the difference-of-convex (DC) property, we transform it into a penalty term and then propose three algorithms, namely the proximal difference-of-convex algorithm (pDCA), pDCA with extrapolation (pDCAe), and the successive convex approximation (SCA), to handle the resulting penalized mean-variance-skewness-kurtosis (PMVSK) formulation. Moreover, we establish theoretical convergence results for 
pDCA and SCA.  Numerical experiments on a real dataset demonstrate the superiority of our proposed methods in obtaining better objective values and sparser solutions efficiently.
\end{abstract}
\begin{keywords}
High-order portfolios, cardinality constraint, difference-of-convex, successive convex approximation
\end{keywords}

\vspace{-0.2cm} 
\section{Introduction}
\vspace{-0.2cm} 
Portfolio management is a fundamental and challenging task for investors. One significant progress was made by Markowitz, who developed the mean-variance (MV) framework~\cite{markowitz19521952}. In the MV framework, the investors' purpose is to maximize their expected profit (i.e., mean return rate, or first moment) and minimize the corresponding risk (i.e., variance of portfolio, or second moment). Due to transaction costs, budget constraints, or even mental costs, investors may pick only a small number of assets out of all possible candidates, thus resulting in the sparse portfolios~\cite{bienstock1996computational}. Some optimization methods have been developed for the MV framework with cardinality constraint ~\cite{bertsimas2018scalable,zhang2019relaxed,frangioni2017improving}. 

The MV framework is based on the assumption that the returns follow a Gaussian distribution or investors have a quadratic utility~\cite{kolm201460}.
However, in real financial markets, the Gaussian distribution assumption is seldom satisfied because asset returns are usually asymmetric and heavy-tailed~\cite{adcock2015skewed,turner1992daily}. To overcome this drawback, several studies took the third and fourth moments (i.e., skewness and kurtosis) into consideration, as they can better capture the asymmetry and heavy-tailed properties of a distribution~\cite{harvey2010portfolio,boudt2015higher,kshatriya2018genetic}. They then proposed the mean-variance-skewness-kurtosis (MVSK) framework, which aims to maximize the mean and skewness while minimizing the variance and kurtosis.
Assuming that investors have constant relative risk aversion (CRRA) \cite{ait2001variable} preferences, the MVSK model is the fourth-order Taylor expansion of the expected utility. Although the MVSK framework is more accurate than the MV framework, it requires solving a non-convex optimization problem, which leads to computational challenges. 

Recently, with the development in the optimization field, some algorithms were proposed to solve the MVSK-related problems. A difference-of-convex algorithm (DCA) was developed in \cite{dinh2011efficient} for the MVSK framework. Further improvement based on the difference-of-convex-sums-of-squares (DC-SOS) was proposed in \cite{niu2019higher}. Quite recently, to approximate the non-convex objective function more tightly, two algorithms based on the majorization-minimization~(MM) and successive convex approximation~(SCA) were proposed in \cite{palomarhopor}. By contrast, the cardinality constrained MVSK (MVSKC) framework has not received much attention yet. The work \cite{brito2019portfolio} considered a bi-objective optimization problem based on the trade-off between expected utility and cardinality. It is shown that there are gains in terms of out-of-sample certainty equivalent and Sharpe ratio for certain cardinality levels. However, they applied a derivative-free solver based on direct multi-search \cite{custodio2011direct}, which may be cumbersome and sensitive to the initial point.

\textbf{Our contributions.} In this paper, based on the available structures of the MVSKC model, we consider integrating the cardinality constraint into the objective function via a penalty technique, which leads to the penalized MVSK (PMVSK) problem. Further, we present three approaches for solving the PMVSK problem, namely proximal difference-of-convex algorithm (pDCA), pDCA with extrapolation (pDCAe), and the successive convex approximation (SCA). Theoretically, we establish the convergence of pDCA and SCA to a stationary point of the problem at hand.
Numerically, we demonstrate through experiments on a real dataset that our methods are able to obtain sparse solutions with lower objective function values. It is worth mentioning that SCA outperforms pDCA and pDCAe in terms of running time, which indicates that it is more favorable for large-scale problems.

\vspace{-0.2cm} 
\section{Problem Formulation}
\vspace{-0.2cm} 
Suppose that the returns of $N$ assets are given by $\Tilde{\boldsymbol{r}}\in \mathbb{R}^N$ and the portfolio weights are given by $\boldsymbol{w}\in \mathbb{R}^N$. The expected return of this portfolio, i.e., the first moment, is
\begin{equation*}
\setlength{\abovedisplayskip}{2pt} 
\setlength{\belowdisplayskip}{2pt}
	\phi_1(\boldsymbol{w}) = E(\boldsymbol{w}^\top\Tilde{\boldsymbol{r}}) = \boldsymbol{w}^\top\boldsymbol{\mu},
\end{equation*}
where $\boldsymbol{\mu} = E(\Tilde{\boldsymbol{r}})$ is the mean return vector. We denote $\boldsymbol{r} = \Tilde{\boldsymbol{r}} - \boldsymbol{\mu}$ as the centered returns. The $q$-th central moment of the portfolio return is $E[(\boldsymbol{w}^\top \boldsymbol{r})^q]$, from which we have the following:
\begin{itemize}
[itemsep= 0.5 pt,topsep = 0.5 pt]
	\item The second moment, a.k.a. variance, of the portfolio return is 
	\begin{equation*}
	\setlength{\abovedisplayskip}{2pt} 
\setlength{\belowdisplayskip}{2pt}
 \phi_2(\boldsymbol{w}) = E[\boldsymbol{w}^\top \boldsymbol{r} \boldsymbol{r}^\top \boldsymbol{w}] = \boldsymbol{w}^\top \boldsymbol{\Sigma}\boldsymbol{w},
	\end{equation*}
	where $\boldsymbol{\Sigma} = E[\boldsymbol{r}\boldsymbol{r}^\top]$ is the co-variance matrix.
	\item The third moment, a.k.a. skewness, of the portfolio
	return is 
	\begin{equation*}
	\setlength{\abovedisplayskip}{2pt} 
    \setlength{\belowdisplayskip}{2pt}
		\phi_3(\boldsymbol{w}) = E[(\boldsymbol{w}^\top \boldsymbol{r})^3] = \boldsymbol{w}^\top \boldsymbol{\Phi} (\boldsymbol{w} \otimes \boldsymbol{w}),
	\end{equation*}
	where $\otimes$ is the Kronecker product and $\boldsymbol{\Phi} = E(\boldsymbol{r}(\boldsymbol{r}^\top \otimes \boldsymbol{r}^\top)) \in \mathbb{R}^{N \times N^2}$ is the co-skewness matrix.
	\item The fourth moment, a.k.a. kurtosis, of the portfolio return is
	\begin{equation*}
	\setlength{\abovedisplayskip}{2pt} 
\setlength{\belowdisplayskip}{2pt}
	\phi_4(\boldsymbol{w}) = E[(\boldsymbol{w}^\top \boldsymbol{r})^4] = \boldsymbol{w}^\top \boldsymbol{\Psi} (\boldsymbol{w}\otimes \boldsymbol{w} \otimes \boldsymbol{w}),
	\end{equation*}
	where $\boldsymbol{\Psi} = E[\boldsymbol{r} (\boldsymbol{r}^\top \otimes \boldsymbol{r}^\top \otimes \boldsymbol{r}^\top)] \in \mathbb{R}^{N\times N^3}$ is the co-kurtosis
	matrix.
\end{itemize}

Based on the above definitions, the MVSKC problem \cite{brito2019portfolio} is given as follows:
\begin{equation}
	\setlength{\abovedisplayskip}{2pt} 
\setlength{\belowdisplayskip}{2pt}
	\begin{split}
		\min_{\boldsymbol{w}} \,&
		f(\boldsymbol{w}) = -\lambda_1 \phi_1(\boldsymbol{w}) + \lambda_2 \phi_2(\boldsymbol{w})
		-\lambda_3 \phi_3(\boldsymbol{w})
		+ \lambda_4 \phi_4(\boldsymbol{w})
		\\
		\mbox{s.t.}\,
		& \boldsymbol{1}^\top \boldsymbol{w}=1,	\|\boldsymbol{w}\|_0 \leq k, -\alpha\boldsymbol{1} \leq \boldsymbol{w} \leq \alpha\boldsymbol{1},
	\end{split}
	\label{eq: mvskcard}
\end{equation}
where $\lambda_1, \lambda_2, \lambda_3, \lambda_4 > 0$ are parameters to balance the four moments of the portfolio return, $\|\boldsymbol{w}\|_0 $ is defined as the number of non-zero elements of $\boldsymbol{w}$, $k<N$ is an integer controlling the number of assets to be selected, and $\alpha > 0$ is to bound each element of $\boldsymbol{w}$.

It is easy to verify that $f$ is non-convex and twice continuously differentiable. The gradient and Hessian of each term in $f$ are given in the following lemma.
\vspace{-0.6\baselineskip}
\begin{lemma}
	\label{lemma: grahess}
	\cite[Lemma 1]{palomarhopor} The gradient and Hessian of the four moments are given by
	\begin{equation*}
\setlength{\abovedisplayskip}{2pt} 
\setlength{\belowdisplayskip}{2pt}
	\begin{split}
		& \nabla \phi_1(\boldsymbol{w}) = \boldsymbol{\mu}, \quad
		\nabla \phi_2(\boldsymbol{w}) = 2\boldsymbol{\Sigma w},\\
		& \nabla \phi_3(\boldsymbol{w}) = 3 \boldsymbol{\Phi} (\boldsymbol{w}\otimes \boldsymbol{w}), \nabla \phi_4(\boldsymbol{w}) = 4 \boldsymbol{\Psi}(\boldsymbol{w}\otimes \boldsymbol{w} \otimes \boldsymbol{w}),\\
		& \nabla^2 \phi_1(\boldsymbol{w}) = \boldsymbol{0}, \quad
		\nabla^2 \phi_2(\boldsymbol{w}) = 2\boldsymbol{\Sigma},\\
		& \nabla^2 \phi_3(\boldsymbol{w}) = 6 \boldsymbol{\Phi}(\boldsymbol{I}\otimes \boldsymbol{w}), \nabla^2 \phi_4(\boldsymbol{w}) = 12 \boldsymbol{\Psi} (\boldsymbol{I}\otimes \boldsymbol{w} \otimes \boldsymbol{w}),
	\end{split}
	\end{equation*}
	where $\boldsymbol{I} \in \mathbb{R}^{N\times N}$ is the identity matrix.
\end{lemma}
\vspace{-0.6\baselineskip}

In addition, we can decompose the objection function $f$ into its convex part $f_{cvx} = -\lambda_1 \phi_1(\boldsymbol{w}) + \lambda_2 \phi_2(\boldsymbol{w})$ and non-convex part $f_{ncvx} =  -\lambda_3 \phi_3(\boldsymbol{w}) + \lambda_4 \phi_4(\boldsymbol{w})$. 
One important observation is that there exists an upper bound on the spectral radius $\rho(\nabla^2 f_{ncvx}(\bm{w}))$ of $\nabla^2 f_{ncvx}(\boldsymbol{w})$, denoted as $\tau_{dc}$ (see Lemma \ref{lemma: rhofbasic}).
Hence, we can rewrite $f(\bm{w})$ as 
\begin{equation*}
\setlength{\abovedisplayskip}{2pt}
\setlength{\belowdisplayskip}{2pt}
f(\boldsymbol{w}) = f_{cvx}(\boldsymbol{w})+ \frac{\tau_{dc}}{2}\boldsymbol{w}^\top \boldsymbol{w} - \left(\frac{\tau_{dc}}{2} \boldsymbol{w}^\top \boldsymbol{w} - f_{ncvx}(\boldsymbol{w})\right),
\end{equation*}
which is a DC function.
\vspace{-0.5\baselineskip}
\begin{lemma}
	\label{lemma: rhofbasic}
	Under the constraints in \eqref{eq: mvskcard}, we obtain 
	\begin{equation}
	\setlength{\abovedisplayskip}{2pt} 
\setlength{\belowdisplayskip}{2pt}
		\label{lemma: rhofncvx}
		\begin{split}
			\ &\rho (\nabla^2 f_{ncvx}(\boldsymbol{w})) \\ 
			\leq 
			\ & 6\alpha\lambda_3 \max_{1\leq i \leq N} \sum_{j=1}^{N^2} |\Phi_{ij}|  
			 + 12 \alpha^2\lambda_4 \max_{1 \leq i \leq N} \sum_{j=1}^{N^3} |\Psi_{ij}|.
		\end{split}
	\end{equation}
\end{lemma}
\vspace{-0.5\baselineskip}

When it comes to the cardinality constraint, inspired by its DC property \cite{gotoh2018dc}
\begin{equation*}
\setlength{\abovedisplayskip}{2pt}
\setlength{\belowdisplayskip}{2pt}
	\| \boldsymbol{w} \|_0 \leq k  \iff \| \boldsymbol{w}\|_1 - \| \boldsymbol{w}\|_{[k]} = 0,
\end{equation*}
where $\| \boldsymbol{w}\|_{[k]}$ is the largest-$k$ norm (i.e., the sum of the $k$ largest elemens in absolute value), 
we transform \eqref{eq: mvskcard} into the following PMVSK problem by taking the cardinality constraint as a penalty term:
\begin{equation}
\setlength{\abovedisplayskip}{2pt}
\setlength{\belowdisplayskip}{2pt}
	\begin{split}
		\min_{\boldsymbol{w}} \quad  f_p(\boldsymbol{w}) = &
	 \dfrac{\tau_{dc}}{2}\boldsymbol{w}^\top \boldsymbol{w} - \left(\dfrac{\tau_{dc}}{2} \boldsymbol{w}^\top \boldsymbol{w} - f_{ncvx}(\boldsymbol{w})\right)  
		\\
		& + f_{cvx}(\boldsymbol{w}) + \rho ( \|\boldsymbol{w}\|_1 -  \| \boldsymbol{w}\|_{[k]})
		\\
		\mbox{s.t.}\quad
		\boldsymbol{1}^\top \boldsymbol{w}=& 1, -\alpha\boldsymbol{1} \leq \boldsymbol{w} \leq \alpha\boldsymbol{1}.
	\end{split}
	\label{eq: mvskcardpenalty}
\end{equation}
Here, $\rho>0$ is the penalty coefficient. 
\vspace{-0.6\baselineskip}

\section{Algorithm Design}
\vspace{-0.2cm} 
\label{sec: algorithm}
In this section, we propose three algorithms---pDCA, pDCAe, and SCA--- to solve the PMVSK problem \eqref{eq: mvskcardpenalty}.
\vspace{-1.0\baselineskip}
\subsection{pDCA}
\vspace{-0.4\baselineskip}
The main idea of pDCA is to successively construct a global upper bound of the objective function in \eqref{eq: mvskcardpenalty} by linearizing the concave part. In the $j$-th iteration, we solve the subproblem
\begin{equation}
\setlength{\abovedisplayskip}{2pt} 
\setlength{\belowdisplayskip}{2pt}
	\begin{split}
		\min_{\boldsymbol{w}} \quad & 
		f_{cvx}(\boldsymbol{w}) + \dfrac{\tau_{dc}}{2}\boldsymbol{w}^\top \boldsymbol{w} + \rho \| \boldsymbol{w} \|_1 \\
		& - (\tau_{dc}\boldsymbol{w}^j-\nabla f_{ncvx}(\boldsymbol{w}^j) + \rho \boldsymbol{s}^j)^\top\boldsymbol{w} 
		\\
		\mbox{s.t.}\quad
		& \boldsymbol{1}^\top \boldsymbol{w}=1,
		-\alpha\boldsymbol{1} \leq \boldsymbol{w} \leq \alpha\boldsymbol{1},
	\end{split}
	\label{eq: mvskcardpdca}
\end{equation}
where $\boldsymbol{s}^j$ is a subgradient of $\| \boldsymbol{w}\|_{[k]}$ at $\bm{w}^j$ and can be computed efficiently through the following two steps: 
\begin{itemize}
[itemsep= 1 pt,topsep = 1 pt]
    \item[1)] sort the elements of $|\boldsymbol{w}^j|$ in decreasing order, i.e., $|w^j_{(1)}|\geq |w^j_{(2)}| \geq \cdots \geq |w^j_{(N)}|$;
    \item[2)] $s^j_{(i)}=\begin{cases}
    \text{sign}(w^j_{(i)}),~~i=1,\dots,k, \\
    0,~~~~\quad \quad \quad{\rm otherwise}.
    \end{cases}$ 
\end{itemize}
%
By introducing a new variable $\boldsymbol{u}\in \mathbb{R}^N$, problem \eqref{eq: mvskcardpdca} can be cast as a convex quadratic programming (QP) problem:
\begin{equation}
\setlength{\abovedisplayskip}{2pt} 
\setlength{\belowdisplayskip}{2pt}
	\begin{split}
		\min_{\boldsymbol{w},\boldsymbol{u}} \quad &
		f_{cvx}(\boldsymbol{w}) + \dfrac{\tau_{dc}}{2}\boldsymbol{w}^\top \boldsymbol{w} + \rho \boldsymbol{1}^\top\boldsymbol{u}  \\
		& - (\tau_{dc}\boldsymbol{w}^j-\nabla f_{ncvx}(\boldsymbol{w}^j) + \rho \boldsymbol{s}^j)^\top\boldsymbol{w} 
		\\
		\mbox{s.t.}\quad
		& \boldsymbol{1}^\top \boldsymbol{w}=1,
		-\alpha\boldsymbol{1} \leq \boldsymbol{w} \leq \alpha\boldsymbol{1},
		-\boldsymbol{u}\leq \boldsymbol{w}\leq \boldsymbol{u},
	\end{split}
	\label{eq: mvskcardpdcaqp}
\end{equation}
which can be efficiently solved by $\mathsf{quadprog}$ in MATLAB. In the rest of the paper, we will always utilize this technique to cast the $\ell_1$ norm into linear inequality constraints. The complete description of pDCA is summarized in Algorithm \ref{alg: pdca}.

\begin{algorithm}[t]
\setlength{\textfloatsep}{0.1cm}
\setlength{\floatsep}{0pt}
	\caption{pDCA for PMVSK \eqref{eq: mvskcardpenalty}.} 
	\label{alg: pdca} 
	\begin{algorithmic}[1] 
		\REQUIRE Iteration number $j$, error tolerance $\epsilon>0$, and initial point $\boldsymbol{w}^0$.
		\ENSURE Optimal solution $\boldsymbol{w}^*$.
		\FOR{$j=0,1,\ldots$}
		\STATE Solve the subproblem \eqref{eq: mvskcardpdcaqp} via QP solver to get $\boldsymbol{w}^{j+1}$.\\
		\IF{$\frac{\|\boldsymbol{w}^{j+1}-\boldsymbol{w}^j \|}{1+ \|\boldsymbol{w}^j\|} < \epsilon$ and $\frac{|f_p(\boldsymbol{w}^{j+1})-f_p(\boldsymbol{w}^{j}) |}{1+ |f_p(\boldsymbol{w}^{j+1})|} < \epsilon$}
		\STATE Set $\boldsymbol{w}^*  = \boldsymbol{w}^{j+1}$.\\
		\PRINT
		\ENDIF
		\ENDFOR
	\end{algorithmic}
\end{algorithm}
\vspace{-1.0\baselineskip}
\subsection{pDCAe}
\vspace{-0.4\baselineskip}
Despite the common use of pDCA in many applications, it can be slow in practice \cite{wen2018proximal}. To accelerate pDCA without increasing too much the computational cost, we adopt an extrapolation technique similar to that in FISTA \cite{beck2009fast} when approximating the concave part of the objective function. In the $j$-th iteration, we solve the subproblem
\begin{equation}
\setlength{\abovedisplayskip}{2pt}
\setlength{\belowdisplayskip}{2pt}
	\begin{split}
		\min_{\boldsymbol{w}} \quad &
		f_{cvx}(\boldsymbol{w}) + \dfrac{\tau_{dc}}{2}\boldsymbol{w}^\top \boldsymbol{w} + \rho \| \boldsymbol{w} \|_1 \\
		& - (\tau_{dc}\boldsymbol{y}^j-\nabla f_{ncvx}(\boldsymbol{y}^j) + \rho \boldsymbol{s}^j)^\top\boldsymbol{w} 
		\\
		\mbox{s.t.}\quad
		& \boldsymbol{1}^\top \boldsymbol{w}=1,
		-\alpha\boldsymbol{1} \leq \boldsymbol{w} \leq \alpha\boldsymbol{1},
	\end{split}
	\label{eq: mvskcardpdcae}
\end{equation}
where $\boldsymbol{s}^j$ is a subgradient of $\| \boldsymbol{w}\|_{[k]}$ at $\boldsymbol{w}^j$. The difference between \eqref{eq: mvskcardpdca} and \eqref{eq: mvskcardpdcae} is that the latter linearly approximates the smooth concave part at $\boldsymbol{y}^j$, which is a carefully designed extrapolation point of $\boldsymbol{w}^{j-1}$ and $\boldsymbol{w}^{j}$. Note that a direct application of the extrapolation technique proposed in \cite{wen2018proximal} for solving \eqref{eq: mvskcardpenalty} can only lead to pDCA due to the nonconvexity of $f$. We summarize the pDCAe in Algorithm \ref{alg: pdcae}.

\vspace{-0.7\baselineskip}
\begin{algorithm}[b]
	\caption{pDCAe for PMVSK \eqref{eq: mvskcardpenalty}.} 
	\label{alg: pdcae} 
	\begin{algorithmic}[1] 
		\REQUIRE Iteration number $j$, error tolerance $\epsilon>0$, and initial values $\boldsymbol{w}^{-1} = \boldsymbol{w}^0$, $\theta_{-1}=\theta_0=1$.
		\ENSURE Optimal solution $\boldsymbol{w}^*$.
		
		\FOR{$j=0,1,\ldots$}
		\STATE Compute $\beta_j = \frac{\theta_{j-1}-1}{\theta_j}$ and $\theta_{j+1} = \frac{1+\sqrt{1+4\theta_j^2}}{2}$.
		\STATE Set $\boldsymbol{y}^j = \boldsymbol{w}^j + \beta_j(\boldsymbol{w}^j - \boldsymbol{w}^{j-1})$. 
		\STATE Solve the subproblem \eqref{eq: mvskcardpdcae} via QP solver to get $\boldsymbol{w}^{j+1}$.
		\IF{$\frac{\|\boldsymbol{w}^{j+1}-\boldsymbol{w}^j \|}{1+ \|\boldsymbol{w}^j\|} < \epsilon$ and $\frac{|f_p(\boldsymbol{w}^{j+1})-f_p(\boldsymbol{w}^{j}) |}{1+ |f_p(\boldsymbol{w}^{j+1})|} < \epsilon$}
		\STATE Set $\boldsymbol{w}^*  = \boldsymbol{w}^{j+1}$.\\
		\PRINT
		\ENDIF
		\ENDFOR
	\end{algorithmic}
\end{algorithm}


\vspace{-0.4\baselineskip}
\subsection{SCA}
\vspace{-0.4\baselineskip}
\begin{algorithm}[t]
	\caption{SCA for PMVSK \eqref{eq: mvskcardpenalty}.} 
	\label{alg: sca} 
	\begin{algorithmic}[1] 
		\REQUIRE Iteration number $j$, error tolerance $\epsilon>0$, and initial point $\boldsymbol{w}^0$.
		\ENSURE Optimal solution $\boldsymbol{w}^*$.
		
		\FOR{$j=0,1,\ldots$}
		\STATE Get $\hat{\boldsymbol{w}}^{j+1}$ by solving a convex QP problem \eqref{eq: scasubproblem}.
		\STATE Perform the backtracking line search \eqref{eq: linesearch} to obtain the stepsize $\gamma^j$.
		\STATE Update $\boldsymbol{w}^{j+1} = \boldsymbol{w}^j + \gamma^j(\hat{\boldsymbol{w}}^{j+1} - \boldsymbol{w}^j)$.
		\IF{$|(\hat{\boldsymbol{w}}^{j+1} - \boldsymbol{w}^j)^\top (\nabla f(\boldsymbol{w}^j) -\rho \boldsymbol{s}^j) + \rho(\| \hat{\boldsymbol{w}}^{j+1} \|_1 - \| \boldsymbol{w}^j \|_1)| < \epsilon$}
		\STATE Set $\boldsymbol{w}^*  = \boldsymbol{w}^{j+1}$.\\
		\PRINT
		\ENDIF
		\ENDFOR
	\end{algorithmic}
\end{algorithm}

In this subsection, instead of utilizing the DC decomposition as in pDCA and pDCAe, we construct a strongly convex function that is not necessarily a global upper bound of $f_{ncvx}(\bm{w})$ but can lead to a tighter approximation. This is a kind of successive convex approximation strategy, and we refer to it as SCA for brevity.

Specifically, we have
\begin{equation*}
\setlength{\abovedisplayskip}{2pt}
\setlength{\belowdisplayskip}{2pt}
\begin{split}
	\Tilde{f}_{ncvx}(\boldsymbol{w},\boldsymbol{w}^j) & = f_{ncvx}(\boldsymbol{w}^j) + \nabla f_{ncvx}(\boldsymbol{w}^j)^\top (\boldsymbol{w} - \boldsymbol{w}^j)  \\
	& \quad+ \frac{1}{2}(\boldsymbol{w}-\boldsymbol{w}^j)^\top \boldsymbol{H}_{ncvx}^j(\boldsymbol{w}-\boldsymbol{w}^j) \\
	& \quad+ \frac{\tau_w}{2}\| \boldsymbol{w} - \boldsymbol{w}^j \|_2^2,
\end{split}
\end{equation*}
where $\boldsymbol{H}_{ncvx}^j \in \mathbb{S}_+^{N}$ is an approximation of $\nabla^2 f_{ncvx}(\boldsymbol{w}^j)$ and $\tau_w>0$ guarantees that $\Tilde{f}_{ncvx}(\boldsymbol{w},\boldsymbol{w}^j)$ is strongly convex.
\vspace{-0.2in}
\begin{lemma}
	\cite[Lemma 5]{higham1988computing} The nearest symmetric positive semidefinite matrix in the Frobenius norm to a real symmetric matrix $\boldsymbol{X}$ is $\boldsymbol{U}\text{Diag}(\boldsymbol{d}_+)\boldsymbol{U}^\top$, where $\boldsymbol{U}\text{Diag}(\boldsymbol{d})\boldsymbol{U}^\top$ is the eigenvalue decomposition of $\boldsymbol{X}$.
\end{lemma}
\vspace{-0.1in}

In addition to constructing a strongly convex local upper bound for $f_{ncvx}(\boldsymbol{w})$, we linearize the concave part $-\rho \| \boldsymbol{w}\|_{[k]}$. In the $j$-th iteration, we solve the following convex QP problem to get $\hat{\boldsymbol{w}}^{j+1}$:
\begin{equation}
\setlength{\abovedisplayskip}{2pt}
\setlength{\belowdisplayskip}{2pt}
	\begin{split}
		\min_{\boldsymbol{w}} \quad &
		f_{cvx}(\boldsymbol{w}) + \rho \|\boldsymbol{w}  \|_1 + \Tilde{f}_{ncvx}(\boldsymbol{w},\boldsymbol{w}^j) - \rho(\boldsymbol{s}^j)^\top \boldsymbol{w}
		\\
		\mbox{s.t.}\quad
		& \boldsymbol{1}^\top \boldsymbol{w}=1, -\alpha\boldsymbol{1} \leq \boldsymbol{w} \leq \alpha\boldsymbol{1}.
	\end{split}
	\label{eq: scasubproblem}
\end{equation}
Note that since problem \eqref{eq: mvskcardpenalty} is non-convex and non-smooth, the traditional choice of stepsize \cite{sun2016distributed} does not work. Hence, we perform a line search to guarantee the convergence of SCA. Specifically, given scalars $0<c<1$ and $0<\beta <1$, the stepsize $\gamma^j$ is set to
be $\gamma^j = \beta^{m_j}$, where $m_j$ is the smallest nonnegative integer $m$
satisfying
\begin{equation}
\setlength{\abovedisplayskip}{2pt}
\setlength{\belowdisplayskip}{2pt}
	\label{eq: linesearch}
	\begin{split}
		&~ f(\boldsymbol{w}^j + \beta^m(\hat{\boldsymbol{w}}^{j+1} - \boldsymbol{w}^j)) - \beta^m \rho (\hat{\boldsymbol{w}}^{j+1} - \boldsymbol{w}^j)^\top\boldsymbol{s}^j \\
		&~~ + \beta^m \rho(\| \hat{\boldsymbol{w}}^{j+1} \|_1 - \| \boldsymbol{w}^j \|_1) \\
	 \leq &~ f(\boldsymbol{w}^j) + c\beta^m (\hat{\boldsymbol{w}}^{j+1} - \boldsymbol{w}^j)^\top (\nabla f(\boldsymbol{w}^j)-\rho \boldsymbol{s}^j) \\
		& ~~+ c\rho \beta^m(\| \hat{\boldsymbol{w}}^{j+1} \|_1 - \| \boldsymbol{w}^j \|_1).
	\end{split}
\end{equation}
It is guaranteed that the stepsize $\gamma^j$ determined by \eqref{eq: linesearch} is non-zero and satisfies $f_p(\boldsymbol{w}^{j+1}) < f_p(\boldsymbol{w}^j)$ \cite[Proposition 2]{yang2018successive}.
The complete SCA algorithm is given in Algorithm \ref{alg: sca}.

\vspace{-0.2cm} 
\section{Theoretical Analysis}
\vspace{-0.2cm} 
\label{sec: theoretical}
In this section, we provide convergence guarantees for our proposed pDCA and SCA in the following theorems.
\vspace{-0.5\baselineskip}
\begin{theorem}
	\label{them: pdcae}
	\cite[Theorem 4.1, Proposition 4.1]{wen2018proximal} Let $\{\boldsymbol{w}^j\}$ be the sequence generated by Algorithm \ref{alg: pdca}. Then the following statements hold:
	\vspace{-0.1cm} 
	\begin{itemize}
		\item $\lim_{j\rightarrow \infty} \|\boldsymbol{w}^{j+1} - \boldsymbol{w}^j \|=0$.
		\item Every limit point of $\{\boldsymbol{w}^j\}$ is a stationary point of \eqref{eq: mvskcardpenalty}.
		\item $f_p^* = \lim_{j\rightarrow \infty} f_p(\boldsymbol{w}^j)$ exists and $f_p \equiv f_p^*$ on $\Omega$, where $\Omega$ is the set of limit points of $\{\boldsymbol{w}^j\}$.
	\end{itemize}
\end{theorem}
\vspace{-0.5cm} 

\begin{theorem}
	\label{them: sca}
	\cite[Theorem 3]{yang2018successive} Let $\{\boldsymbol{w}^j\}$ be the sequence generated by Algorithm \ref{alg: sca}. Then, every limit point of $\{\boldsymbol{w}^j\}$ is a stationary point of \eqref{eq: mvskcardpenalty}.
\end{theorem}

\vspace{-0.6cm} 
\section{Experiments}
\vspace{-0.2cm}

\label{sec: exp}
In this section, we evaluate the performance of our proposed algorithms in a real dataset. The data are generated according to the following steps:
\vspace{-0.2cm} 
\begin{enumerate}
	\item[1)] Randomly select $N$ ($N=50$) stocks from S$\&$P 500 Index components.
	\item[2)] Randomly choose $5N$ continuous trading days from {\tt 2012-12-01} to {\tt 2018-12-01}.
	\item[3)] Compute the sample moments using selected data.
\end{enumerate}
\vspace{-0.1cm} 

Our experiments are performed in MATLAB on a PC with i5-6200U CPU at 2.3GHz and 12GB memory. We set $\alpha = 0.2$, $k=10$, $\epsilon = 10^{-8}$, $\tau_w= 10^{-10}$, $\rho = 4\times 10^{-3}$ for pDCA, pDCAe, and SCA. In addition, we set model parameters in $f$ following \cite{boudt2015higher} with $\lambda_1=1$, $\lambda_2=\xi/2$, $\lambda_3=\xi(\xi+1)/6$, and $\lambda_4=\xi(\xi+1)(\xi+2)/24$, where $\xi$ is the risk aversion parameter.

\begin{figure}[btp]
\setlength{\abovecaptionskip}{-0.3cm} 
    \centering
    \subfigure[]{
    \includegraphics[width=0.23\textwidth]{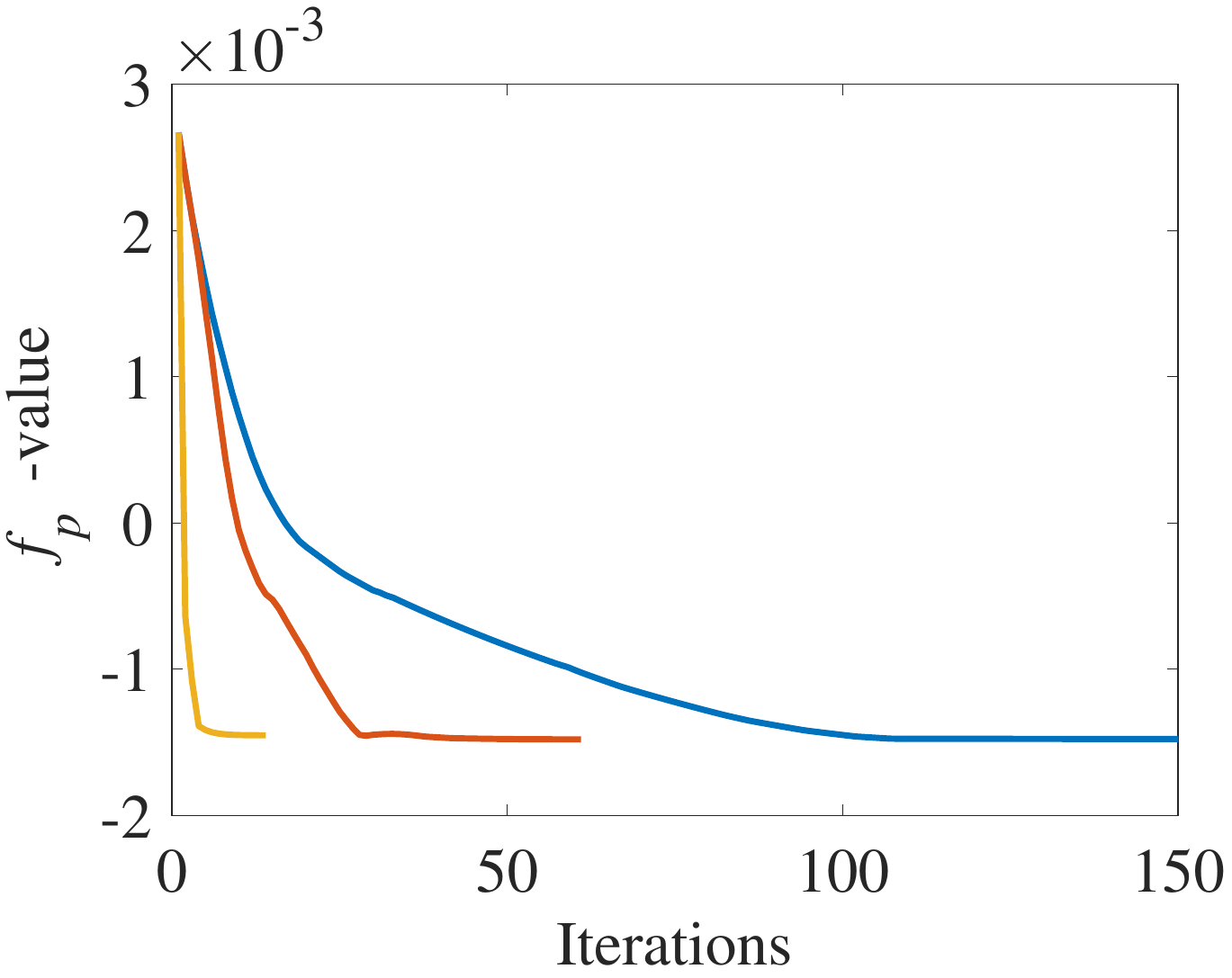}}
    \subfigure[]{
    \includegraphics[width=0.22\textwidth]{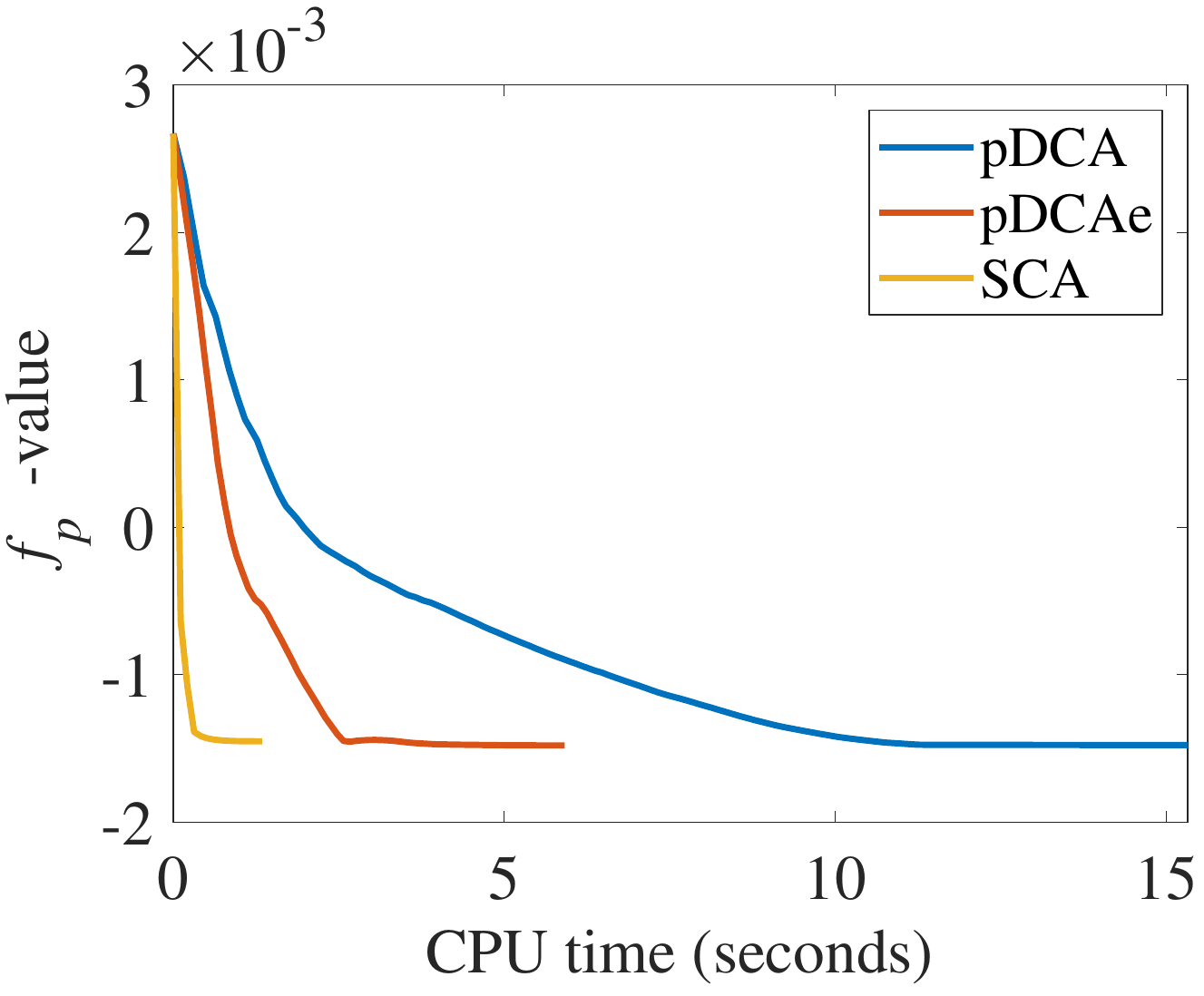}}
    \vspace{-0.15in}
    \caption{$f_p$-value versus iteration/CPU time with $\xi=10$.}
    \label{Fig.main_1}
    \vspace{-0.8\baselineskip}
\end{figure}

\begin{figure}[btp]
\setlength{\abovecaptionskip}{-0.2cm} 
\setlength{\belowcaptionskip}{-2cm}
    \centering
    \subfigure[]{
    \includegraphics[width=0.23\textwidth]{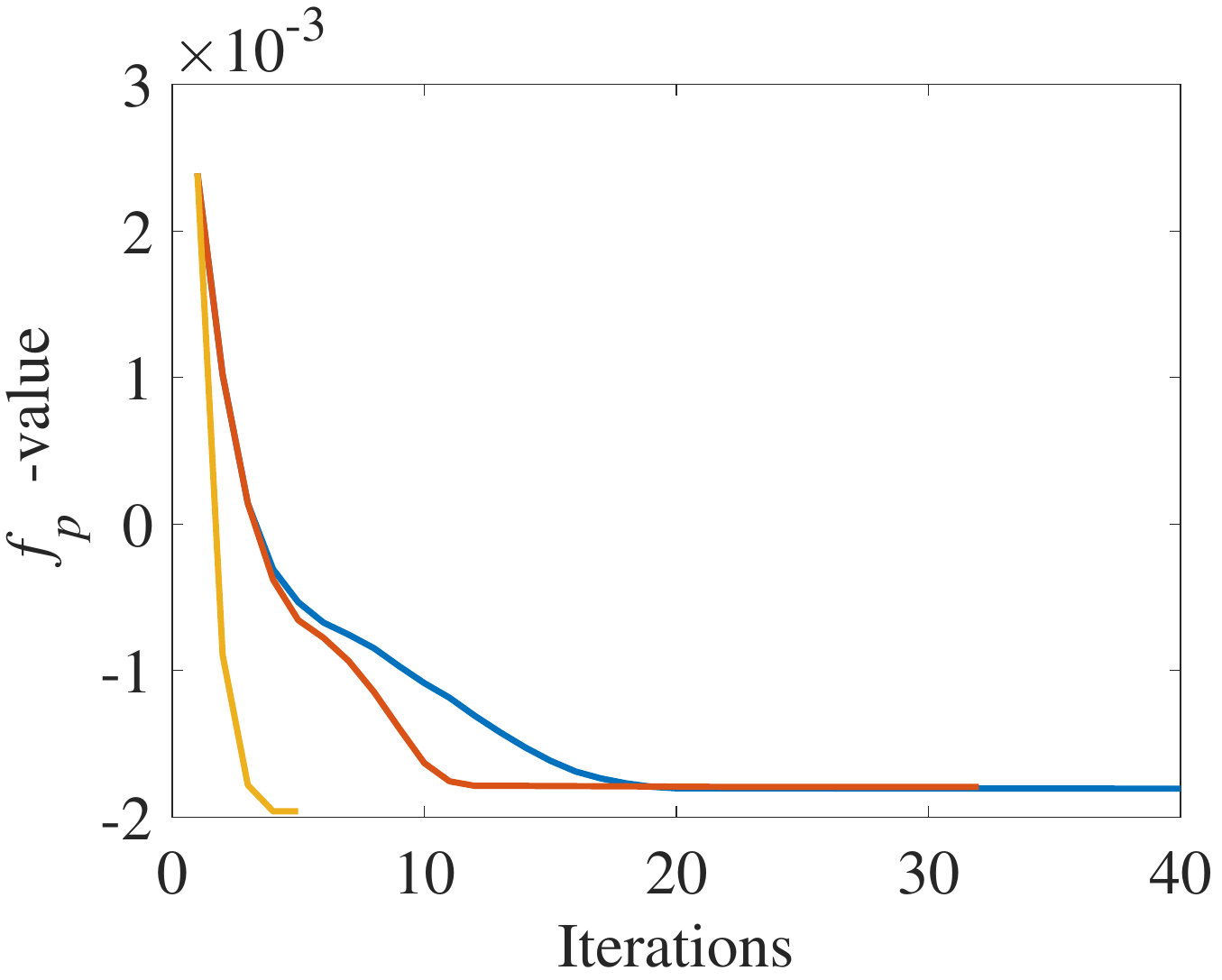}}
    \subfigure[]{
    \includegraphics[width=0.22\textwidth]{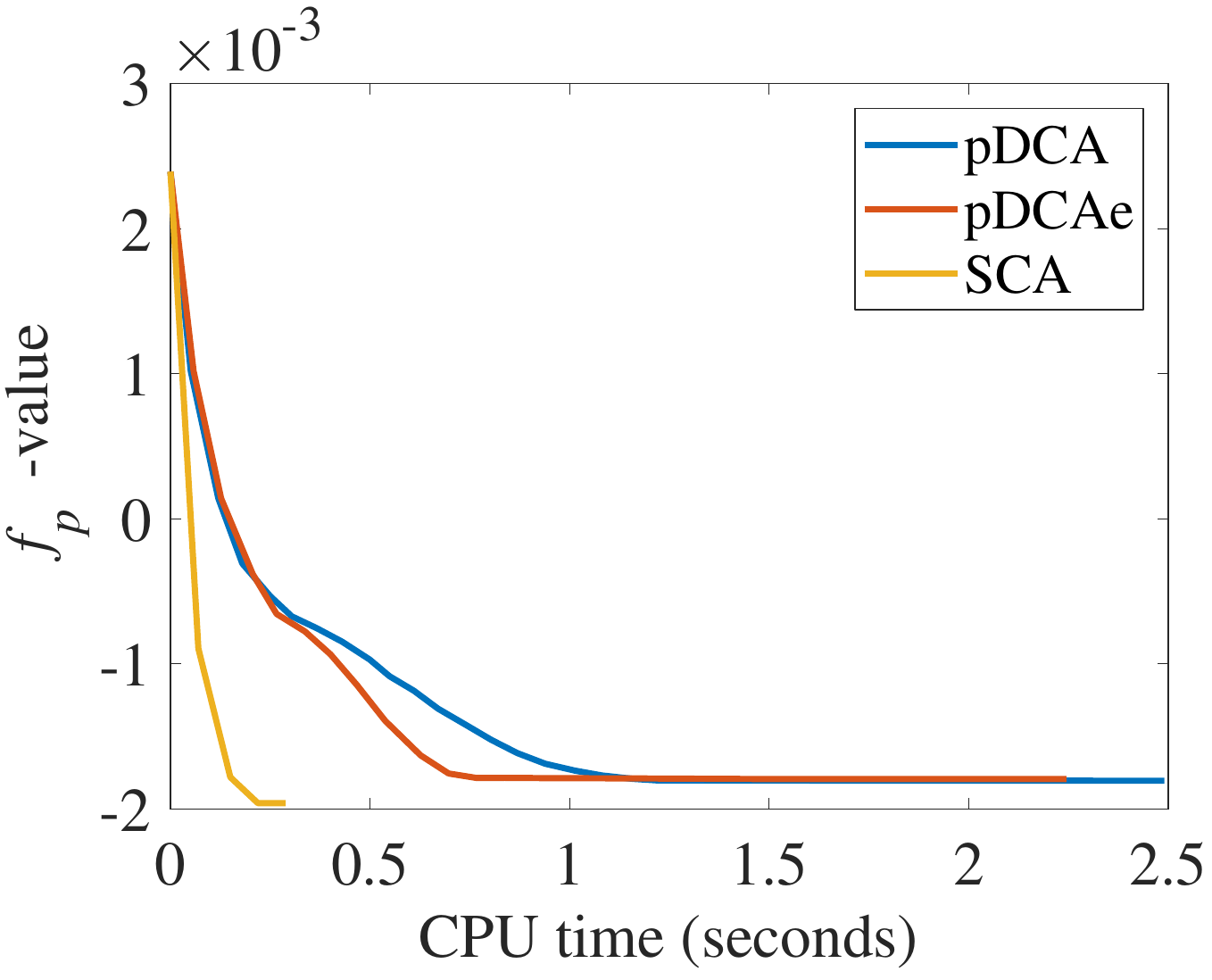}}
    \vspace{-0.15in}
    \caption{$f_p$-value versus iteration/CPU time with $\xi=5$.}
    \label{Fig.main_2}
\vspace{-0.9\baselineskip}
\end{figure}

We first show the convergence curves w.r.t. the iteration number and CPU time. Fig. \ref{Fig.main_1} shows the objective value curves of pDCA, pDCAe, and SCA with $\xi=10$ and the same random initialization. Fig. \ref{Fig.main_2} presents the corresponding results when $\xi=5$. Here, we only show the first 150 and 40 iterations of the algorithms for $\xi=10$ and $\xi=5$,  respectively. We can observe from Fig. \ref{Fig.main_1} (a) and Fig. \ref{Fig.main_2} (a) that SCA converges fastest among the three algorithms, and pDCAe converges faster than pDCA. Since the QP subproblems~\eqref{eq: mvskcardpdcaqp}--\eqref{eq: scasubproblem}, which constitute the main computational burden of the corresponding algorithms, require roughly the same cost, fewer iterations indicate less running time. This is verified in Fig. \ref{Fig.main_1} (b) and Fig. \ref{Fig.main_2} (b), which show that SCA takes the least time among the three methods.

Next, we compare our proposed algorithms with several baselines. One commonly used strategy, which we denote as relaxed MVSKC (RMVSKC), is to first relax the cardinality constraint to $\ell_1$ norm constraint \cite{palomarhopor} (it is worth emphasizing that this kind of relaxation is not able to get a sparse solution) and then project the resulting solution $\Tilde{\boldsymbol{w}}^*$ to satisfy the original constraints of MVSKC \eqref{eq: mvskcard}. The projection step amounts to solving 
\begin{equation*}
\setlength{\abovedisplayskip}{2pt}
\setlength{\belowdisplayskip}{2pt}
	\begin{split}	\min_{\boldsymbol{w}} \quad&
		\frac{1}{2}\| \boldsymbol{w} - \Tilde{\boldsymbol{w}}^*\|^2
		\\
		\mbox{s.t.}\quad
		& \boldsymbol{1}^\top \boldsymbol{w}=1,	\|\boldsymbol{w}\|_0 \leq k, -\alpha\boldsymbol{1} \leq \boldsymbol{w} \leq \alpha\boldsymbol{1}.
	\end{split}
	\label{eq: mvskl1project}
\end{equation*}
To this end, we introduce a binary variable $\boldsymbol{u}$ to rewrite the cardinality constraint, leading to $-\alpha \boldsymbol{u} \leq \boldsymbol{w} \leq \alpha \boldsymbol{u}$, and $\boldsymbol{1}^\top \boldsymbol{u} \leq k$. However, the resulting mixed-integer quadratic programming problem is challenging even with Gurobi (e.g.,  it takes more than $5$ hours using Gurobi with $\Tilde{\boldsymbol{w}}^*\in\mathbb{R}^{50}$). Hence, we use the genetic algorithm (GA) instead. Utilizing the same transformation technique, we can also solve the MVSKC \eqref{eq: mvskcard} via GA. From Table \ref{table:1}, we can see that pDCA, pDCAe, and SCA are superior in terms of the objective values obtained. In addition, it is achieved with much less time when using SCA.
\vspace{-0.05in}

\vspace{-0.2in}
\begin{table}[hbtp] 
    \caption{$f$ value and time usage of different methods with $\xi=10$ and $\xi=5$. Each method is tested five times using different initializations and the results are then averaged.}
    \centering
	\setlength{\tabcolsep}{1.08mm}{
    \begin{tabular}{c|cc||cc}
    \hline
    \multirow{2}{*}{Methods} & \multicolumn{2}{c||}{$\xi=10$}&\multicolumn{2}{c}{ $\xi=5$} 
    \cr\cline{2-5}
    & $f$ value & CPU time (s) & $f$ value & CPU time (s) \\
    \hline
    pDCA & -1.40e-3 & 19.5 & -2.00e-3 & 11.6\\
    \hline
    pDCAe & -1.60e-3 & 6.4 & -2.00e-3 & 2.6 \\
    \hline
    SCA & -1.10e-3 & \textbf{0.7} & -1.70e-3 & \textbf{0.7} \\
    \hline
    RMVSK & +9.00e-4 & 9.4 & +4.00e-4 & 309.1 \\
    \hline
    MVSKC & +8.00e-4 & 80.3 & +4.00e-4 & 92.0\\
    \hline
    \end{tabular}
    }
    \label{table:1}
\end{table}


\vspace{-1.5\baselineskip}
\section{Conclusion}
\vspace{-0.7\baselineskip}
\label{sec: conclusion}
In this paper, we considered the problem of high-order portfolio optimization with cardinality constraint. We proposed to recast the cardinality constraint into a penalized term and then developed three methods, namely pDCA, pDCAe, and SCA. Convergence results were established. Extensive experiments showed that our methods
got lower objective values and are more efficient than the baselines.
\vfill\pagebreak

\bibliographystyle{IEEEbib}
\bibliography{refs}

\end{document}